\documentclass[aoas,preprint]{imsart}

\usepackage{color}
\usepackage{bm,verbatim }
\usepackage{graphicx}
\usepackage{subfigure}
\usepackage{amsfonts,amssymb,latexsym}
\usepackage[mathscr]{eucal}

\RequirePackage[OT1]{fontenc}
\RequirePackage{amsthm,amsmath}
\RequirePackage{natbib}

\startlocaldefs
\numberwithin{equation}{section}
\theoremstyle{plain}

\begin{document}

\begin{frontmatter}

\title{Assignment of endogenous retrovirus integration sites using a mixture model}
\runtitle{Mixture Model for Retrovirus Integrations}

\author{\fnms{David R.} \snm{Hunter$^1$}\ead[label=e2]{dhunter@stat.psu.edu}}
\and
\author{\fnms{Le} \snm{Bao$^1$}\ead[label=e1]{lebao@psu.edu}}
\and
\author{\fnms{Mary} \snm{Poss$^2$}\ead[label=e3]{mposs@bx.psu.edu}}
\affiliation{Department of Statistics$^1$, Penn State University, University Park, PA, USA\\
Department of Biology$^2$, Penn State University, University Park, PA, USA\\
\printead{e1,e2,e3}}

\runauthor{D.~Hunter et al.}

\affiliation{Pennsylvania State University}

\maketitle

\begin{abstract}
Structural variation occurs in the genomes of individuals because of the different positions occupied by 
repetitive genome elements like endogenous retroviruses, or ERVs.  The presence or absence of 
ERVs can be determined by identifying the junction with the host genome using 
high-throughput sequence technology and a clustering algorithm. 
The resulting data give the number of sequence reads assigned to 
each ERV-host junction sequence for each sampled individual.  Variability in the number of
reads from an individual integration site makes it difficult to determine whether a site is present
for low read counts.
We present a novel two-component mixture of negative binomial distributions to model these counts and
assign a probability that a given ERV is present in a given individual.
We explain how our approach is superior to existing alternatives, including
another form of two-component mixture model and the much more common approach of selecting
a threshold count for declaring the presence of an ERV.
We apply our method to a data set of ERV integrations in mule deer [Odocoileus hemionus], a species for which no 
genomic resources are available, and demonstrate that the discovered patterns of shared integration sites contain 
information about animal relatedness.

\end{abstract} 


\end{frontmatter}

\section{Introduction}
\label{sec:introduction}

Determining how genome sequences vary among individuals and populations is an important
research area because genetic differences can confer phenotypic differences. The most commonly
reported variations in genome sequence between two individuals are those that occur at the
nucleotide level, e.g,  single nucleotide polymorphisms (SNPs).  These are typically identified 
by comparing the nucleotide at each position of a query sequence to that of a reference genome. 
Individual genomes can also differ in the relative position and number of homologous genome 
regions. For example, a genetic locus can be duplicated, deleted, inverted or moved to a new 
location in one genome compared to another. These changes in the genome are called genome 
structural variations (GSVs) and are more difficult to analyze than SNPs, particularly if a region is 
present in the query but absent from the reference.  Transposable elements (TEs) are an important 
type of GSV that comprise over 50\% of most eukaryote genomes \citep{Cordaux2009}. 
TEs are capable of moving in the genome by several mechanisms, including a copy-paste 
mechanism \citep{Kazazian2004}. 
Although many TEs are fixed in the genome of a species---that is, 
all individuals will have the TE at a specific 
location in the genome---others are present in 
some individuals and absent in others, which results in 
polymorphism at the site of the TE insertion.

Because TEs have important phenotypic consequences on the host genome \citep{Kazazian2004,
Bohne2008,Bourque2009,Odonnell2010,Fedoroff2012,Kapusta2013,Kokosar2013}, 
it is important to have robust methods to determine the location of a specific element in genomes
so as to know if the element is present or absent from an individual.
These data can be obtained by molecular approaches that amplify the region spanning the end of 
the TE and the adjacent genome region of the host; a product is obtained only if the TE is present. 
Multiple methods have been developed to detect different classes of TEs in the genomes of 
individuals via high throughput sequencing \citep{Odonnell2010,Iskow2010}, 
allowing investigators to identify the location of all TEs of a specific type in an individual 
genome.  Yet even in a well-annotated genome like the human genome, new mobile elements
are sometimes discovered in poorly annotated regions \citep{Contreras-Galindo2013}.
For this reason, mapping the sequence reads representing the sites of element integration
to a reference genome is insufficient even in well-annotated genomes. 
Furthermore, in many species, no genome exists or, if it does, the completeness is much less than 
for humans.  Indeed, in the case that we consider here, no genome exists for any 
member of the cervidae.

\citet{Bao2014} 
reported recently on a method to detect an endogenous retrovirus (ERV), which is a type of TE 
derived from an infectious retrovirus, in the genome of the Cervid mule deer (Odocoileus hemionus), 
a species that lacks a reference genome. Each Cervid endogenous 
retrovirus (CrERV) 
is present at a unique position in the genome \citep{Elleder2012,Wittekindt2010}---which we refer to
as an ``insertion site'' throughout this article---and because the infections giving rise to the 
CrERVs are relatively recent, the prevalence of individual CrERVs can vary from a single animal to a majority 
of the population. Animals that share an insertion site must be related because once acquired,
ERVs are inherited along family lineages like any host gene.  Thus, animals with similar profiles of 
CrERV insertion sites in their genomes share an ancestral lineage and have the potential to display similar 
phenotypic effects of CrERV compared to animals without CrERVs at those sites. In order to 
investigate the consequences of CrERV integration on the mule deer host, 
\citet{malhotra2016} developed a de novo 
clustering approach that groups all insertion sites that occupy the same genomic region from different animals. 
Each cluster of sequences may be represented by a single consensus sequence that in turn represents 
the site in the host genome where the virus has integrated.  The resulting data are 
an $m\times n$ matrix $X$, where 
the $(i,j)$ element $X_{ij}$ gives the count of sequences (the read count) 
from animal $j$ that are assigned to CrERV-host
junction $i$, which will henceforth be referred to as insertion site $i$.  
Here, $m$ and $n$ are the total numbers of sites and animals, respectively.

\begin{figure}[htb]
\includegraphics[width=.8\textwidth]{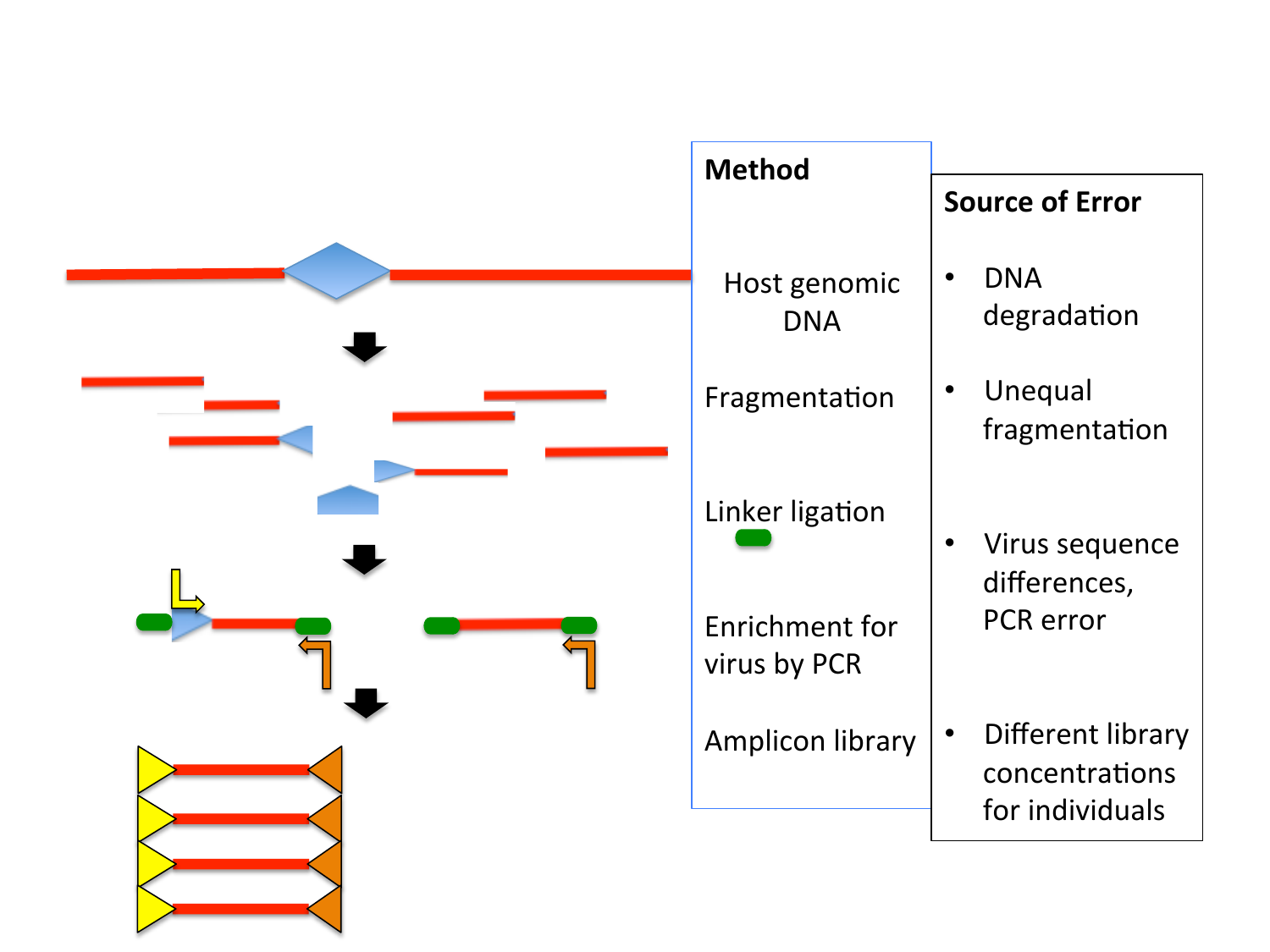}
\caption{\footnotesize{
Schematic diagram showing the steps used to generate the junction fragment libraries and 
some of the sources of error leading to variation in read count data.
Red lines indicate host genomic DNA and the blue diamond represents the site of a mobile 
element insertion---in our case, a retrovirus.  
The right hand box gives some of the reasons that read counts vary among animals and replicates.
Additional potential sources of error include 
unequal mixing, barcode contamination \citep{faircloth2012}, and sequencing error,
which can affect clustering.}}
\label{fig:methods}
\end{figure}

These read count data contain information about 
whether an individual carries specific integration sites. However, 
read counts may contain both false positives and false negatives: A small number of sequences 
may be attributed to an animal not carrying a particular insertion site due to either 
measurement errors in the high-throughput methods or mis-assignment in the 
clustering process; and no sequences may be captured for an animal actually carrying 
a particular site when there are insufficient sequences.
Some of the sources of error leading to highly variable read counts for an 
integration site are shown in Figure~\ref{fig:methods}, 
which also gives an overview of the data pipeline.  That figure illustrates that 
the DNA (red lines) is fragmented, fragments are selected for 
size compatible with the sequencing platform (typically 300--500 bp), and small DNA 
oligonucleotides (linkers, in green) are ligated to the ends.  
The fragments containing the mobile element 
are enriched by polymerase chain reaction or PCR, employing a primer specific to the 3' portion of the 
retrovirus and one in the linker, which yields a product containing the sequence of the host-retroviral 
junction.  The linkers are engineered so that the primer cannot bind until the virus-specific primer 
has first generated a strand of DNA; thus, if the virus is not present, there is no amplification of 
the fragment.  Individual libraries from different samples are mixed together in equal molar 
amounts after being tagged by library-specific DNA ``barcode'' sequences, and all libraries 
are sequenced together.  

The problem of accuracy of low read counts is well known for high-throughput sequence data,
as can be seen in the report by \citet{baillie2011} and the subsequent reanalyses of the data by 
\citet{evrony2012} and \citet{evrony2016}
that documented many false positives associated with inclusion of low read count data.
It is therefore challenging to determine the true status of insertion site $i$ in animal $j$ when read counts are low.
One approach is to set a threshold, and assume that a site is 
carried by an animal whenever the corresponding read count is above the threshold. 
This ad-hoc practice has serious drawbacks, as discussed in \citet{Bao2014}:  Essentially, it ignores
differences in the genomic integration sites, some of which are more readily sequenced than others; 
quality of the DNA; laboratory error; and sequence quality, which varies between sequencing runs.
Any of these factors can cause wide variation in total read number per animal and per integration site.
Although \citet{Bao2014} move beyond the naive thresholding approach 
by proposing a mixture model, the mixture used in that article
of a Poisson component and a truncated geometric
component has several drawbacks. 
The present article presents a much-improved mixture model
that attempts to account for these sources of variability.
We then describe the reasons for modeling choices and discuss 
the results of fitting this model to the read count data.

We have made the following materials available as a supplementary {\tt .zip} file:
The original (unabridged) dataset of read counts obtained by the clustering algorithm, along with the
abridged version used for the analyses in this article; the code, written for the R computing 
environment \citep{r2016}, 
that reproduces all of the analyses, tables, and figures in this article; and the additional datasets
used for the plotting the latitude/longitude coordinates of the animals as well as the 
wet-bench data obtained using PCR
for ground-truthing a subset of the classification results obtained by various models.

\section{A mixture model approach}
\label{mixmodapproach}

Count data are sometimes modeled using a Poisson distribution or, if more flexibility is required,
a negative binomial distribution.  When in addition some of the counts are zeros created by
a separate random mechanism, we may introduce a point mass at zero; the resulting
``zero-inflated'' count models are in fact simplistic mixture models.
For our data, zero-inflation is not sufficient because even nonzero counts $X_{ij}$
may occur when insertion site $i$ is absent from animal $j$.  
Instead, we must account for counts both when site $i$ is present and when it is absent.  The counts
that are observed in each of
these two cases will be modeled as one component of our two-component mixture model.  Our goal
in developing this model is to respect model parsimony as well as the experimental realities of
the sequencing processes used to obtain the data.  This section explains our modeling
choices and, in particular, why we have avoided the mixture of Poisson and 
truncated geometric distributions originally used by \citet{Bao2014}, which appears to be the only 
previous mixture-model-based treatment in the literature of this type of count data.

\subsection{Improving the Mixture Model}
\label{sec:mixmod}
\begin{figure}[htb]
\begin{tabular}{cc}
\begin{minipage}{.45\textwidth}
\includegraphics[width=\textwidth]{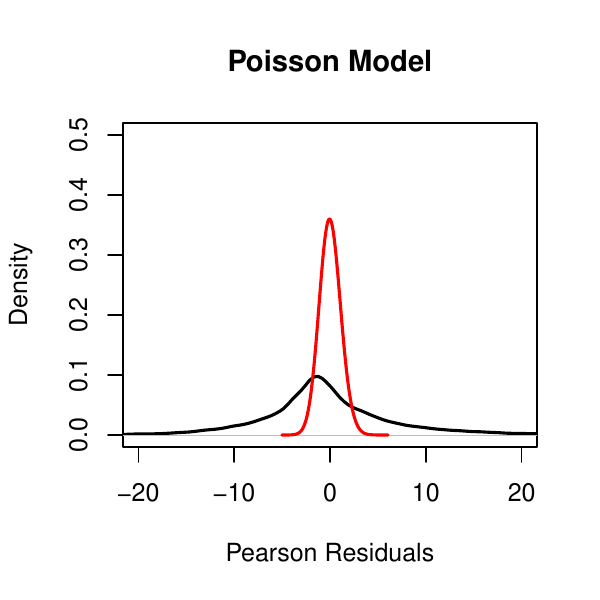}
\end{minipage}
&
\begin{minipage}{.45\textwidth}
\includegraphics[width=\textwidth]{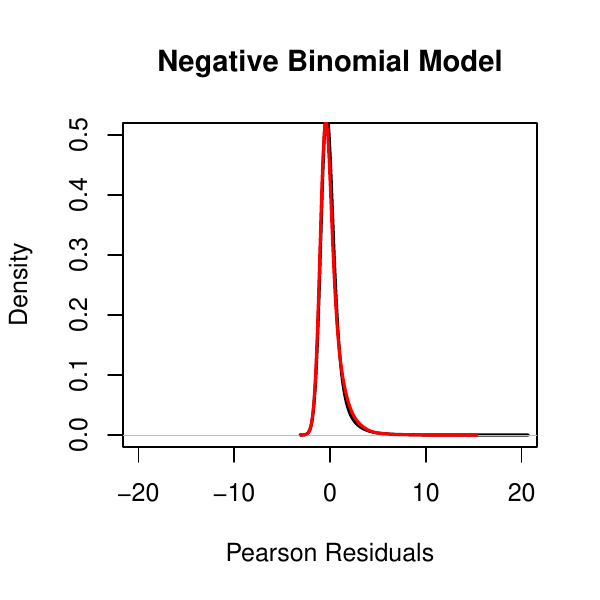}
\end{minipage}
\\
\begin{minipage}{.45\textwidth}
\includegraphics[width=\textwidth]{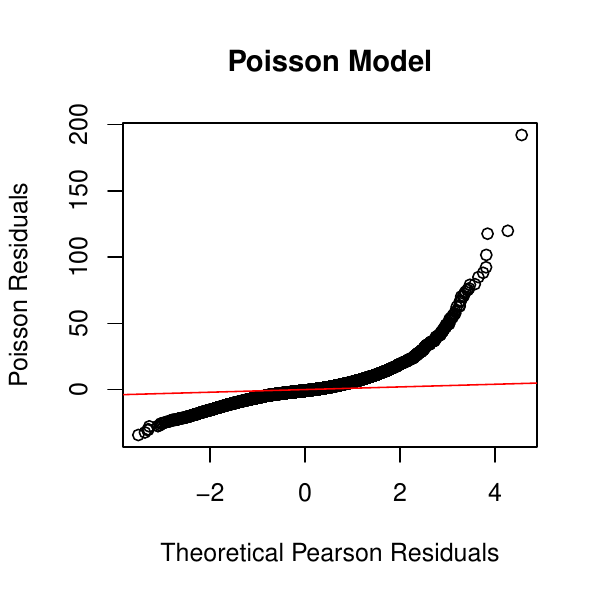}
\end{minipage}
&
\begin{minipage}{.45\textwidth}
\includegraphics[width=\textwidth]{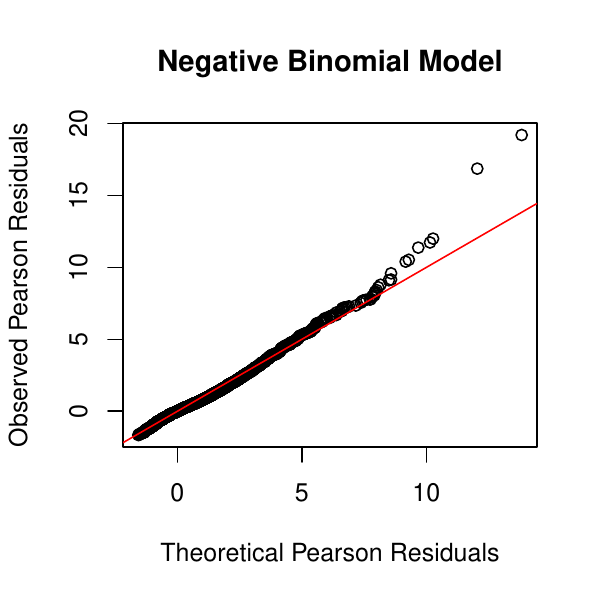}
\end{minipage}
\\
\end{tabular}
\caption{\footnotesize{
The top two panels show the kernel density estimates of the Pearson residuals
(solid lines) and their theoretical densities (dashed lines) for 
all counts with probability at least 0.5 of being from the ``present''
mixture component according to the mixture model.  The lower panels are Q-Q plots of the 
observed vs. theoretical Pearson residuals.}}
\label{Over-dispersion}
\end{figure}

Let us first consider the situation in which animal $j$ carries insertion site $i$, which we call
the ``present'' case because site $i$ is truly present, as opposed to the ``absent'' case where
it is not.  
The first model that springs to mind for count data is something based
on the Poisson distribution; indeed, this is the approach used by
\cite{Bao2014}.  However, we have found strong evidence of over-dispersion---that is, 
evidence that the standard deviation of these present counts is larger than the square root
of their mean---even when we use a model with a large number of parameters to
account for the heterogeneities across animals and insertion sites.  
This over-dispersion is depicted in Figure~\ref{Over-dispersion}, 
which compares the best-fitting Poisson and 
negative binomial models in terms of their Pearson residuals, which are the observed 
counts minus the estimated counts divided by the square roots of the estimated variances.
On the other hand, the negative binomial family appears adequate for this
modeling task.  

To help explain how the plots in Figure~\ref{Over-dispersion} were created, we first introduce 
both the Poisson and negative binomial models for the ``present'' mixture component.  
In the Poisson case, ``present'' counts $X_{ij}$ are assumed to be distributed independently
as Poisson$(a_i b_j)$ for parameters $a_1, \ldots, a_m$ and $b_1, \ldots, b_n$.
Thus, the probability mass function for $x=0, 1, 2, \ldots$ is
\begin{equation}\label{poisson}
P(X_{ij}=x) = \exp\{-a_i b_j\} \frac{ (a_i b_j)^{x_{ij}} } {x_{ij}!},
\end{equation}
and $E(X_{ij}) = \mbox{Var}(X_{ij}) = a_i b_j$.
In the negative binomial plots, the assumption is that the $X_{ij}$ are distributed independently
as negative binomial random variables with parameters $r_j$ and $\alpha_i$ for
$1\le i\le m$ and $1 \le j\le n$.  This gives
\begin{equation}
\label{truePositive}
P(X_{ij}=x) = f_{ij}(x; r, \alpha) \stackrel{\rm def}{=}
\binom{x+r_j-1}{x} \alpha_i^{r_j} (1-\alpha_i)^{x},
\end{equation}
a mass function with mean 
$E(X_{ij}) = r_j(1-\alpha_i)/\alpha_i$ and 
variance $\mbox{Var}(X_{ij}) = r_j(1-\alpha_i)/\alpha_i^2$.

The plots in the left column of Figure~\ref{Over-dispersion} are obtained by
fitting the counts $X_{ij}$ using 
a slightly improved version of the model used by \citet{Bao2014}, namely, a
two-component mixture model
where one component (``present'') is the Poisson distribution of Equation
(\ref{poisson}) and the other component (``absent'') is a truncated geometric distribution.
In addition to maximum likelihood estimates of the $a_i$ and $b_j$ parameters,
the fitting procedure yields estimates of the conditional probabilities of inclusion in the
``present'' component for each $X_{ij}$ observation.  In creating the plots, 
we consider only those $X_{ij}$ with
estimated probabilities greater than $0.5$ in constructing the plots, which is a simplistic
way to focus on the fit of only the ``present'' (Poisson) component.  The Pearson residuals
are calculated for these $X_{ij}$ by subtracting the corresponding estimated mean
$\hat a_i \hat b_j$ and dividing by the estimated standard deviation 
$(\hat a_i \hat b_j)^{1/2}$.  
Whenever an animal is replicated in the dataset---that is, whenever there exist two labels 
$j'\ne j$ for the same
animal---we constrain the model so that the estimates of the ``present'' probability must be equal.
This is different from the model used by \citet{Bao2014}, which treated replicates as independent
samples for the purposes of estimation.
Finally, the theoretical distribution that forms the basis
of comparison for the plots is obtained via simulation from the distribution determined by 
the fitted parameters.  The plots in the right column of Figure~\ref{Over-dispersion}
are obtained in the same way, except that the mixture model uses the 
negative binomial distribution of Equation~(\ref{truePositive}) for the ``present'' component
and another negative binomial distribution for the ``absent'' component.
Further discussion about our choice for the ``absent'' component is provided below.

Based on Figure~\ref{Over-dispersion}, 
the data clearly suggest discarding the Poisson model in favor of the negative binomial model
for the ``present'' mixture component of the model.  Interestingly, this choice is not merely
in favor of the model with more parameters, as is often the case when a negative binomial
distribution fits better than a Poisson distribution; here, each model has the same number of parameters.  
In equation~(\ref{truePositive}),
we interpret $\alpha_i$ as an insertion site-specific parameter where
$1-\alpha_i$ approximates the enrichment of site $i$, and $r_j$ as an animal-specific
parameter.
Thus, the mean and variance are both directly proportional to the animal-specific $r_j$ parameter
and they are decreasing functions of the site-specific $\alpha_i$ parameter.

\begin{figure}[htb]
\includegraphics[width=3in]{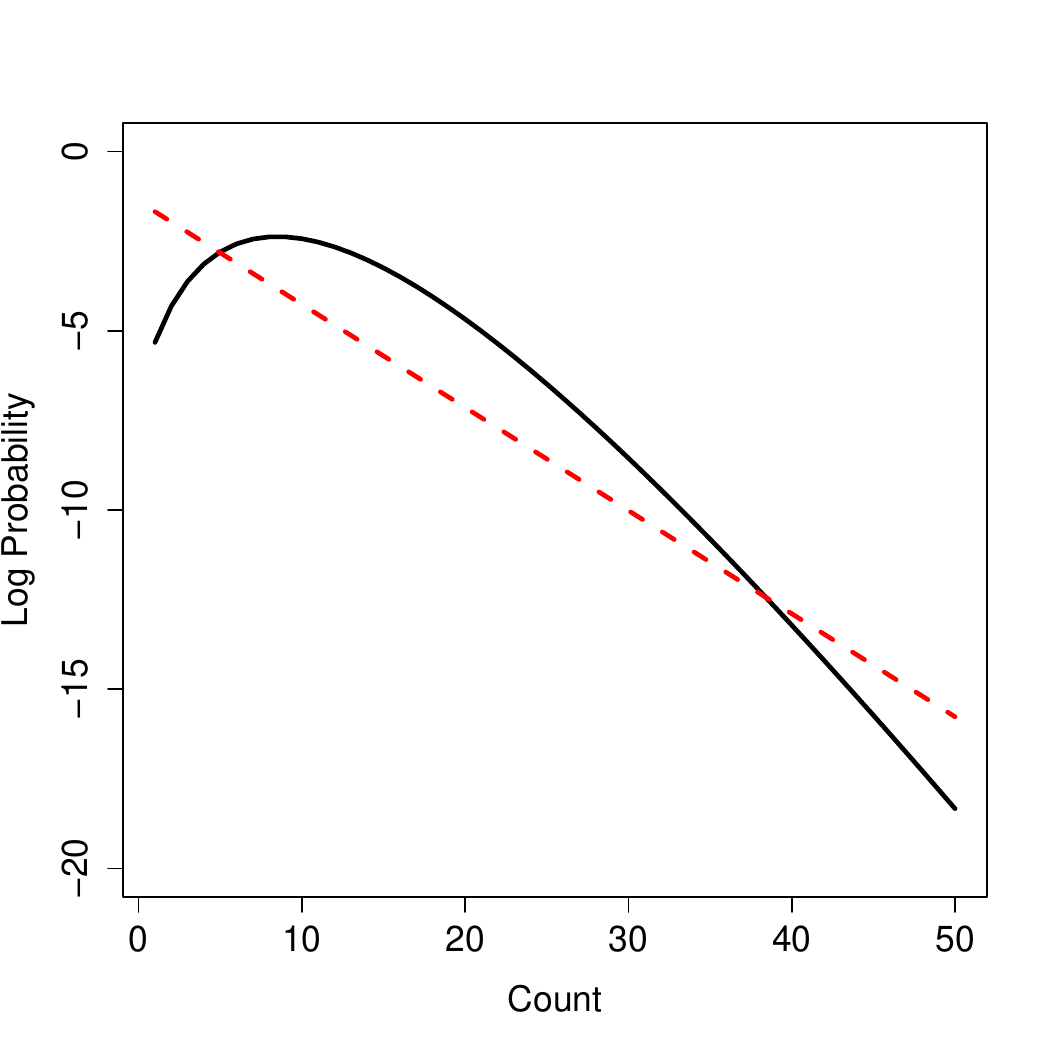}
\caption{The two mass functions whose logarithms are shown here are a negative
binomial with parameters $r=10$ and $\alpha=1/2$ (solid line) and a geometric
with parameter $p=1/4$ (dashed line).  The means of the distributions are 10 and 3, 
respectively, even though the geometric has a larger mass function for large values.}
\label{geomVsNegBinom}
\end{figure}

In the ``absent'' case 
where animal $j$ does not carry insertion site $i$, in principle we may choose an entirely 
different class of distributions to model the observed counts.  We reject the 
class of Poisson 
distributions immediately 
because we need a distribution with a variance
substantially larger than its mean.  
The geometric distribution is an interesting potential alternative and has 
the advantage of simplicity since, like the Poisson, it only requires a single parameter.  
However, we reject the geometric
for a different reason:  The geometric mass function decays more slowly for large values 
than that of a negative binomial,
even if the mean of the former is smaller than the mean 
of the latter, as illustrated in Figure~\ref{geomVsNegBinom}.  
Thus, outlying large counts could be classified as coming from the absent component,
which is nonsensical.  \citet{Bao2014}
sidestepped this issue by
truncating the geometric distribution of counts from the absent component.  However, we wish to 
avoid the problematic question of how to choose a truncation point.

Due to the considerations above, we reject both the Poisson and geometric models for counts from the
absent mixture component in favor of a more flexible negative binomial model
and posit that whenever animal $j$ does not carry insertion site $i$, the mass function for the count 
$X_{ij}$ is given by
\begin{equation}
\label{trueNegative}
g_{ij}(x; r, p) \stackrel{\rm def}{=} \binom{x+r_j-1}{x} p_{k(j)}^{r_j} (1-p_{k(j)})^{x}, \quad x=0, 1, 2 ,\ldots
\end{equation}
where $k(j)$ is the batch in which animal $j$ was sequenced, $r_j$ is the same 
animal-specific parameter as in Equation~(\ref{truePositive}),
and the expected false-positive count for the $k(j)$
batch is a decreasing function of $p_{k(j)}$:  As explained earlier, 
this expected count is $r_j (1-p_{k(j)}) / p_{k(j)}$.

Both negative binomial distributions, in Equation~(\ref{truePositive}) and Equation~(\ref{trueNegative}),
can be interpreted 
as a sum of $r_j$ independent geometric distributions.  This is a deliberate modeling choice that reflects the
fact that the quality and quantity of each
animal's sample may vary, and this variation will affect 
counts from both the present class and the absent class in the same way.

The insertion site-specific effect is most relevant in the present case, as reflected by the fact that we allow
counts from the present component to depend on the parameter $\alpha_i$ where $i$ denotes the site number.
In the absent case, counts may be considered to be ``background noise'' and 
therefore likely to depend on the particular batch but not the insertion site in question; for this reason,
we allow the count distribution for the absent class 
to depend on $p_{k(j)}$, where $k(j)$ denotes the batch number of animal $j$.  

We occasionally obtain distinct sets of counts from the same animal when samples from the
same animal are run in different batches.  In such a case, our model 
treats these counts as though they come from different animals, conditional on the mixture component
from which the counts $X_{ij}$ come.  That is, each set of counts
receives its own index $j$, so the $r_j$ parameters may be different.  This is important since 
different sets of counts come from distinct batches, and these often have dramatically different
count profiles.  In fact, allowing for this flexibility, which is enhanced by indexing the 
absent count distributions by $p_{k(j)}$, means that our model can easily accommodate new data as they are 
created in separate sequencing runs or on separate sequencing platforms.  
Accommodating new data 
is scientifically important, since our data are continually updated as new animals
are sequenced; sequencing technology advances rapidly, and it is not always feasible
nor cost-effective to rerun previously sequenced animals using newer technology.  Thus, our
method allows for seamless data integration by preventing us from having to discard 
useful data simply because technology changes or our set of sampled animals expands.

On the other hand, it is important that our model can account for cases in which
multiple sets of counts come from the same animal
in our dataset.  This is
done by placing appropriate constraints on the mixing probabilities $\pi_{ij}$, where 
$\pi_{ij}$ represents the a priori probability that animal $j$ carries insertion site $i$.  Thus,  we introduce the constraint $\pi_{ij}=\pi_{ij'}$ for any $j\ne j'$ for which $j$ and $j'$
index the sets of counts from two different runs on the same animal. 
Once we introduce the $\pi_{ij}$ probabilities, the full likelihood of our mixture model becomes
\begin{equation}
\label{likelihood}
L( \pi, r, \alpha, p) =
\prod_{i=1}^m \prod_{j\in U} \left[ \pi_{ij} \prod_{j' \in S_j} f_{ij'}(x; r, \alpha) + (1-\pi_{ij}) \prod_{j'\in S_j} g_{ij'}(x; r, p)  \right],
\end{equation}
subject to the constraints explained above,
where $S_j = \{ j' : \mbox{$j$ and $j'$ are the same animal}\}$ and $U$ is 
any set containing exactly one element from each $S_j$;
that is, $U$ is a set of indices for the unique animals.
We experimented with three simple parameterizations
of the $\pi_{ij}$ parameters:  (1) $\pi_{ij}=\pi$ for all $i$ and $j$; (2) $\pi_{ij}=\pi_i$ for all $j$; and 
(3) $\pi_{ij}=\pi_j$ for all $i$.  
It may be surprising that, say, parameterization (1) is at all interesting;
however, the question of whether animal $j$ truly includes insertion site $i$---which clearly
depends on $i$---is different from the question of whether the {\em proportion} of such inclusions depends on $i$.
The latter is an empirical question that should be examined using the data.  
In Section~\ref{sec:estimates}, 
we find that option (2) attains the best Akaike's Information Criterion (AIC) score.

Since Equations~(\ref{truePositive}) and~(\ref{trueNegative})
represent the present and absent components, respectively, it seems reasonable
to conjecture that $E(X_{ij})$ will be 
greater in Equation~(\ref{truePositive}) than in Equation~(\ref{trueNegative}).
These two means are given by
$r_j(1- \alpha_i)/\alpha_i$ and $r_j (1-p_{k(j)})/p_{k(j)}$, respectively.  Thus, 
since the $r_j$ parameter is common to the two mass functions,
the conjectured inequality may
be guaranteed by enforcing the constraints $\alpha_i<p_{k}$ for all $i$ and $k$ during
the estimation procedure.  
Enforcing such constraints in an algorithm, say, by always updating $\alpha_i$ as the
smaller of the ECM algorithm estimate and the minimum current $p_k$ value,
would not in principle complicate the computations.
However, such enforcement
would potentially destroy the ascent property of the algorithm mentioned in 
Section~\ref{sec:parameterEstimation} and, perhaps more importantly,
it would raise the troubling question of how to 
interpret final parameter estimates for which some $\alpha_i=p_k$.  
If we instead choose to
enforce some positive gap $g$ between the largest $\alpha_i$ and the smallest $p_k$,
then we are faced with the arbitrary choice of a value of $g$.
We therefore opt not to enforce such constraints, yet 
we find nevertheless that our unconstrained point estimates satisfy $\alpha_i<p_{k}$.
The fact that we obtain these results without enforcing them is an encouraging sign for the model fit.
We discuss the actual estimated values in Section~\ref{sec:estimates}.

\subsection{Parameter Estimation}
\label{sec:parameterEstimation}

Estimation of the model parameters is accomplished using maximum likelihood via a straightforward 
Expectation-Conditional Maximization (ECM) algorithm \citep{Meng1993}.  Essentially, an ECM algorithm
is merely an EM algorithm in which only one subset of the parameters is updated at each iteration or sub-iteration.  
The goal is to maximize the log-likelihood function of the parameters $\pi$, $r$, $\alpha$, and $p$.  This goal is 
complicated by the fact that we do not observe which data come from the first mixture component and which come
from the second.  Typically, one approaches this problem by defining indicator variables
\[
Z_{ij} = I\{\mbox{animal $i$ carries insertion site $j$} \};
\]
these $Z_{ij}$ are then considered missing, or unobserved, data, and an EM algorithm aims to maximize the
log-likelihood based on only the observed data by exploiting the mathematically simpler form of the log-likelihood
based on the full data in a clever way, alternating between an E-step and an M-step.

In the E-step, given the iteration-$t$ parameter values
$\pi_{ij}^{(t)}$, $r_j^{(t)}$, $\alpha_i^{(t)}$, and 
$p_{k(j)}^{(t)}$, we calculate the probability that animal $i$ carries insertion site $j$:
\begin{equation}\label{originalZ}
Z_{ij}^{(t)} = \frac{\pi_{ij}^{(t)} \prod_{j'\in S_j} f_{ij'}^{(t)}(x)}{\pi_{ij}^{(t)} \prod_{j'\in S_j} f_{ij'}^{(t)}(x) 
+ (1-\pi_{ij}^{(t)}) \prod_{j'\in S_j} g_{ij'}^{(t)}(x)}.
\end{equation}
In the M-step (actually the CM-step), 
we update the parameters in four distinct subsets, in each case holding the other parameters fixed at their most
up-to-date values.  To wit, we first consider the $\alpha$ parameters.  We find that
the log-likelihood involving $\alpha_i$ is
\[
\sum_{i,j}Z_{ij}^{(t)} \left[ \log \binom{x_{ij}+r_j^{(t)} -1}{x_{ij}}+ r_j^{(t)} \log(\alpha_i) + x_{ij} \log (1-\alpha_i)   \right],
\]
which is maximized at 
\[
{\alpha}_i^{(t+1)} = \frac{\sum_j Z_{ij}^{(t)} r_j^{(t)}}{\sum_j Z_{ij}^{(t)} \left(x_{ij} + r_j^{(t)} \right)}.
\]
The estimate of $\alpha_i$ will be unstable if $Z_{ij}$ is close to zero for all $j$, so in practice, we let 
\[
{\alpha}_i^{(t+1)} = \frac{\sum_j Z_{ij}^{(t)} r_j^{(t)}+0.05}{\sum_j Z_{ij}^{(t)} \left(x_{ij} + r_j^{(t)} \right)+0.1},
\]
noting that the ascent property guaranteed by an ECM algorithm relies only on the assurance that the complete-data log likelihood 
increases its value at each iteration; if the corrected version of $\alpha^{(t+1)}$ ever fails to produce such an increase, it may be
replaced by the exact version.

The log-likelihood that involves $p_{k(j)}$ is maximized at 
\[
{p}_k^{(t+1)} = \frac{\sum_{j \in A_k} \left(1-Z_{ij}^{(t)} \right) 
r_j^{(t)}}{\sum_{j \in A_k} \left( 1-Z_{ij}^{(t)} \right ) \left(x_{ij} + r_j^{(t)} \right)},
\]
where $A_k$ denotes the set of animals coming from the $k$th batch.
The log-likelihood that involves $r_j$ is maximized at
\[
\begin{array}{ccl}
{r}_j^{(t+1)} 
&=&  \arg \max_{r_j} \Bigl\{ 
\sum_i \log \binom{x_{ij}+r_j -1}{x_{ij}} \\
&& +  r_j \sum_i \left[ Z_{ij}^{(t)}\log \alpha_i^{(t+1)}   
+  \left(1-Z_{ij}^{(t)} \right) \log p_{k(j)}^{(t+1)}  \right] \Bigr\},
\end{array}
\]
which will be solved numerically.  Finally, there are several different update formulas for the $\pi_{ij}$ parameters, 
depending on which of the three models we are using.  We have
\[
\pi^{(t+1)} = 
\frac{1}{mn} \sum_{i=1}^m \sum_{j=1}^n Z_{ij}^{(t)}, \ 
\pi_i^{(t+1)} = 
\frac{1}{n} \sum_{j=1}^n Z_{ij}^{(t)}, \mbox{\ or\ }
\pi_j^{(t+1)} = 
\frac{1}{m} \sum_{i=1}^m Z_{ij}^{(t)},
\]
depending on the whether we select model (1), (2), or (3), respectively.

We initialize the ECM algorithm at $Z_{ij}^{(0)}=\max(1, x_{ij}/10)$ and $r_j^{(0)} = 100$, 
and estimate parameters by iterating between the M-step and the E-Step described above.

We stop iterating when the sum of the absolute changes of all $Z_{ij}^{(t)}$ is less than 0.01;
these values at convergence will be denoted by $\hat Z_{ij}$; they represent the probabilities,
conditional on the observed data, that animal $j$ has insertion site $i$ when the parameter values
are taken to be the maximum likelihood estimates.  The $m\times n$ matrix of all such probabilities
will be denoted $\hat Z$.

Because EM-based algorithms can be sensitive to starting parameter values, we also explore different starting values. 
Letting $Z_{ij}^{(0)} =\max(1, x_{ij}/c)$ where $c=2,3,\ldots,20$, and letting $r_j^{(0)}$ vary from 5 to 500, we
find that all these combinations of starting values converge to essentially the same solution.

After the algorithm has converged, the entries of the matrix $\hat Z$ may be used as
estimates of the probabilistic assignment of insertion sites to animals, which may in turn lead
to insights into the relationships among animals.  We revisit this topic in Section~\ref{sec:animals}.

\section{Results}
\label{sec:result} 
The $1722\times 77$ matrix $X$ containing the read count data is provided in the supplementary materials
in the file {\tt ReadCount.csv}.  
This dataset is an abridged
version of the original dataset, which is entitled {\tt UnabridgedReadCount.csv}, 
that excludes any insertion sites that do not have at least two samples containing more than five 
reads.  The reason for this choice is to focus on only rows that provide substantial evidence 
for relatedness of two or more animals; however, our model can in principle easily handle rows with 
low read counts.
Of all the read counts in the abridged table, 82.6\% are zero and another 6.3\% are between one 
and ten, inclusive.  The mean of all non-zero counts is 98.6.

\subsection{Mixture model parameter estimates}
\label{sec:estimates}

Section~\ref{sec:mixmod} presents three models for the $\pi_{ij}$ parameters, 
and  we use ``an information theoretic criterion,'' also known as Akaike's information criterion or AIC 
\citep{akaike1974}, to select from among them.  The results are displayed in Table~\ref{AICscores}.
While there is no consensus about which of several possible model selection 
criteria based on penalized log-likelihood scores 
should be used in the context of mixture models, in this case the differences among the three 
likelihoods are so large that the choice is clear regardless of the criterion we use.
\begin{table}[htb]
\begin{tabular}{| l | c | cc |} \hline
&Number of & \multicolumn{2}{|c|}{Treatment of replicates} \\ \cline{3-4}
\qquad Model & model parameters & Independent samples & Identical animals \\ \hline
(1) $\pi_{ij}=\pi$ & 1803 &$346{,}013$ & 324,245 \\
(2) $\pi_{ij}=\pi_i$ & 3524 & ${\bf 318{,}015}$ & ${\bf 301{,}951}$ \\
(3) $\pi_{ij}=\pi_j$ & 1879 & $344{,}276$ & $324{,}246$ \\ \hline
\end{tabular}
\caption{Akaike's information criterion (AIC) scores, 
given by $-2$ times the maximized log-likelihood plus $2d$, where $d$ is the number of
model parameters.}\label{AICscores}
\end{table}
Model (2) is selected as the best model regardless of whether we consider re-tested samples
from the same animal as independent samples or not.  This model assumes that 
the prevalence rates of insertion sites are 
heterogeneous, and the following analysis focuses on the $\pi_{ij}=\pi_i$ setting.  

Our primary interest is in the matrix $\hat Z$, which is a $1722\times 77$ matrix, and in addition to these
values there are 3524 parameter estimates.  Estimates of the sampling distributions of these
parameters based on asymptotic normality are unlikely to be productive since the huge 
$3524\times 3524$ covariance matrix will be impossible to estimate
given only $1722\times 77$ data points, and even with sparsity 
constraints the estimation would be difficult.
A parametric bootstrap approach is possible whereby we use the estimated parameters to repeatedly
simulate new entire count datasets, obtain sets of bootstrap parameter estimates for each one, 
and take the empirical multivariate distribution of the bootstrap estimates to estimate 
the sampling distributions of the 
original estimates.  An interesting question of implementation arises as to whether we should
treat the $Z_{ij}$ as parameters or data:  The former idea suggests that we should simulate
site assignment indicator variables 
according to the estimated $\hat Z_{ij}$ values, whereas the latter suggests that we should
simulate site assignments using only the $\hat \pi_i$ estimates.  To give a sense of the possibilities,
we have included R code in the supplemental files
that implements the former idea---though the code may be easily modified
to implement the latter---and yields standard errors 
for the estimates $\hat p_1=0.967$, $\hat p_2=0.939$, and $\hat p_3=0.980$ 
of $0.005$, $0.014$, and $0.001$, respectively, based on 500 bootstrap samples.  
We may conclude for instance that our
three experiments yield statistically significantly different false positive rates $1-p_k$.
However, we do not undertake here a thorough exploration of the use of the bootstrap
in this context.

\begin{figure}[htb]
\begin{tabular}{cc}
\begin{minipage}{.45\textwidth}
\includegraphics[width=\textwidth]{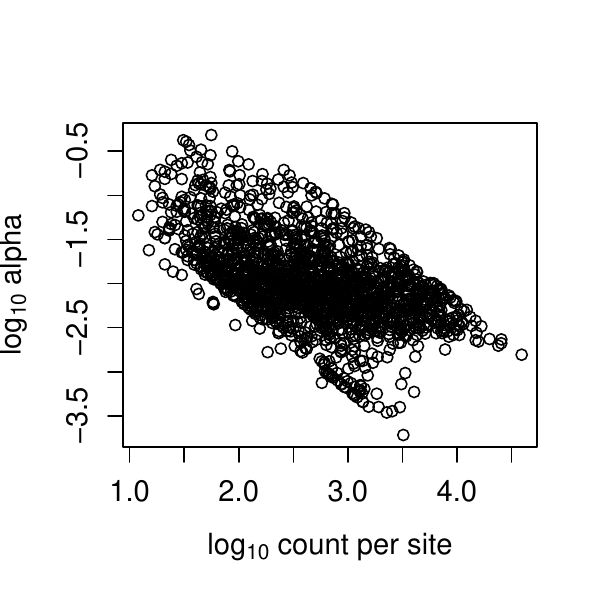}
\end{minipage}
&
\begin{minipage}{.45\textwidth}
\includegraphics[width=\textwidth]{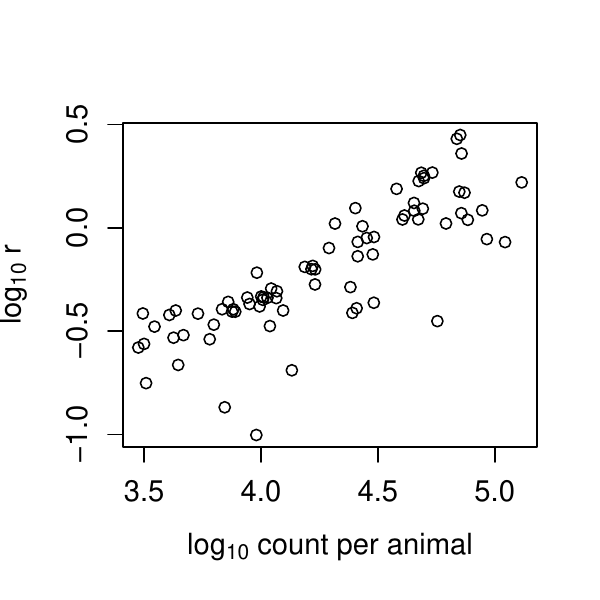}
\end{minipage}
\\
\begin{minipage}{.45\textwidth}
\includegraphics[width=\textwidth]{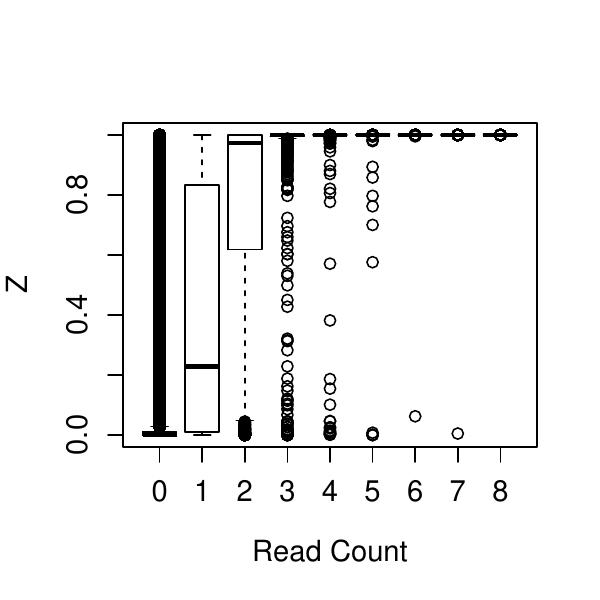}
\end{minipage}
&
\begin{minipage}{.45\textwidth}
\includegraphics[width=\textwidth]{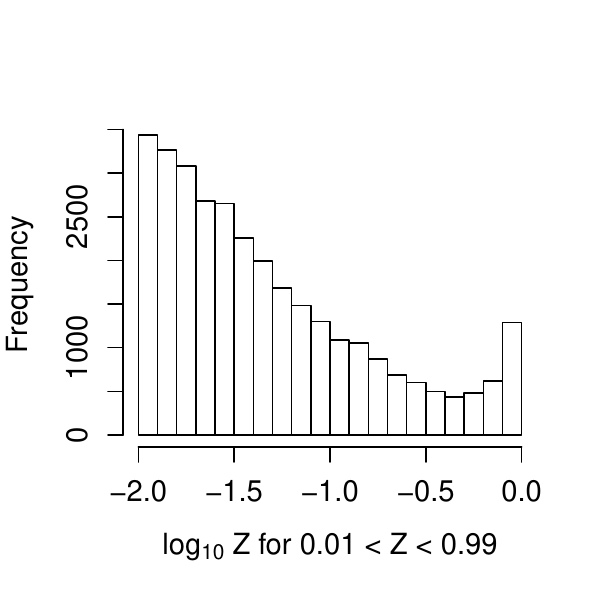}
\end{minipage}
\\
\end{tabular}
\caption{\footnotesize{
Top row:  Sums of row counts plotted against corresponding $\alpha_i$ and $r_j$
parameter estimates.
Bottom left:  Boxplots of $\hat Z_{ij}$ as a function of read count for counts up to 8.  
Bottom right:  Histogram of $\log_{10} \hat Z_{ij}$ values for $0.01\le\hat Z\le0.99$.  We obtain
61\% of $\hat Z_{ij}$ values less than .01 and 15\% greater than 0.99.
}}
\label{fig:estimates}
\end{figure}

For the model that takes replicated animals into account, 
the estimates of the 1722 $\alpha_i$ parameters range from 
$1.92\times 10^{-4}$ to $0.483$, and 
the estimates of the $p_{k(j)}$ parameters are 0.967, 0.939, and 0.980.  We see therefore that
$\alpha_i<p_{k(j)}$ in each case, which
is sensible as we noted in Section~\ref{sec:mixmod}, 
even though we do not enforce this constraint in the 
optimization algorithm.  This result guarantees that the expected read counts for the 
``present'' case are always larger than those 
for the ``absent'' case.  

Figure~\ref{fig:estimates} depicts some characteristics of the $\alpha_i$, $r_j$, 
and $Z_{ij}$ estimates.
The corresponding results for the model treating the replicated animals as independent
are similar graphically, so we omit them here.
The top plots show a roughly log-log relationship between total count and the corresponding
$\alpha_i$ or $r_j$ parameter.
The bottom left plot shows that
the estimated $\hat Z_{ij}$ values are not a monotone function of the 
read counts $X_{ij}$, 
which demonstrates that the mixture model approach 
captures subtle individual heterogeneities among
insertion sites and animals that a simplistic threshold cannot.  
We see that some $\hat Z_{ij}$ values near zero correspond 
to counts at least as large as those corresponding to some $\hat Z_{ij}$ values near one.

\begin{table}[htb]
\begin{tabular}{|c|cc|cc|cc|c|} \hline
  & \multicolumn{2}{|c|}{Observed}  & \multicolumn{2}{|c|}{Independent\ NB} & \multicolumn{2}{|c|}{Independent\ PTG} & Replicate \\ 
Site  & \multicolumn{2}{|c|}{Counts}  & \multicolumn{2}{|c|}{Probabilities} & \multicolumn{2}{|c|}{Probabilities} & NB and PTG\\ \cline{2-7}
(Cluster) & batch S & batch M & batch S & batch M & batch S & batch M & Probabilities \\ \hline
Cluster107  &  0 & 498 & 0.20 & 1.00 & 0.00 & 1.00 & 1.00 \\
Cluster436  &  0  & 84 & 0.08 & 1.00 & 0.00 & 1.00 & 1.00 \\
Cluster315  &  2  & 72 & 0.99 & 1.00 & 0.00 & 1.00 & 1.00 \\
Cluster296  & 44   & 0 & 1.00 & 0.04 & 1.00 & 0.00 & 1.00 \\
Cluster591  &  0  & 42 & 0.01 & 1.00 & 0.00 & 1.00 & 1.00 \\
Cluster403  & 42   & 1 & 1.00 & 0.05 & 1.00 & 0.00 & 1.00 \\
Cluster166  & 41   & 1 & 1.00 & 0.75 & 1.00 & 0.00 & 1.00 \\
Cluster62    & 0  & 37 & 0.53 & 1.00 & 0.00 & 1.00 & 1.00 \\
Cluster199  &  2  & 35 & 1.00 & 1.00 & 0.00 & 1.00 & 1.00 \\
Cluster1729 &  0  & 33 & 0.00 & 1.00 & 0.00 & 1.00 & 1.00 \\ \hline
\end{tabular}
\caption{Probabilities of ``present'' 
for various models for 10 examples of extremely divergent counts for the same 
animal/site combination, all taken from animal 01.  Here, ``NB'' stands for 
the negative binomial model of Equations~(\ref{truePositive}) and~(\ref{trueNegative}),
whereas ``PTG'' stands for the Poisson-truncated geometric model of \citet{Bao2014}.
}
\label{deer01counts}
\end{table}

Also interesting is the effect on certain $\hat Z_{ij}$ values of the requirement that samples from the
same animal must have the same estimated probabilities of inclusion in the ``present'' 
mixture component.  Because there is so much variability present in the read counts,
it is not uncommon to find wildly different counts for the same animal at the same 
insertion site.  As an example, let us consider animal 01 from the dataset, which 
occurs in two different batches, labeled the S batch and the M batch.  
Table~\ref{deer01counts} gives the ten largest read counts for this animal that are paired
with counts not greater than 2.  When such cases arise in the model that considers replicate
information, we see that they are essentially classified with probability one as ``present''---that is,
a single large count is sufficient for such a classification even in the presence of a low count in a 
different batch.  
In other words, a single large count appears sufficient to categorize an animal into the ``present'' 
component, and when the same animal/site combination also yields small counts, the model
can adjust both by allowing for a large variance and because each sample has a unique $r_j$
parameter even if it comes from an animal with more than one sample.
On the other hand, for the model that considers all samples as independent, a low 
count can result in a small estimate of the ``present'' probability 
even though the presence or absence of a given insertion site
in a given animal should not change from batch to batch.

\subsection{Ground-truthing various models}
\label{sec:validation}

It is possible to verify the presence or absence of a particular insertion site in a particular animal
by directly visualizing the DNA fragment amplified at a particular integration site using PCR.
In this way, as in \citet{Bao2014}, we have obtained the true status of 
6 insertion sites in 32 unique animals, as summarized in the supplemental file
{\tt PCRVerificationData.csv}.
Some of these animals occur in more than one batch, so we have
a total of 45 samples to consider, comprising a total of 
$6\times 45 = 270$ probability assignments to the ``present'' mixture component.  
Some of these probabilities will be constrained to be equal when we consider models
that take replications of animals into consideration.

As a measure of model performance, we report area under the receiver operating characteristic curve, 
or AUC \citep{bradley1997}, 
which in this case may be calculated as follows:  First, we note that 47 of the 270 probability assignments 
correspond to truly present integration sites, whereas the remaining 223 correspond to absent sites.
We examine all $47\times 233$ pairs of discordant pairs, and calculate the proportion of these pairs
in which the estimated probability for the truly present site exceeds the estimated probability for the
absent site.  For purposes of this calculation, cases in which the estimated probabilities are equal 
for a discordant pair are counted as one-half.
The results of this analysis are presented in Table~\ref{AUCtable}.
In addition to four different statistical models, we also consider the read counts themselves, which
may be subjected to the same AUC calculation.  We may view this model-free analysis an
upper bound on the potential performance of any possible thresholding procedure, since such 
a procedure uses a fixed read count as the cutoff between ``absent'' and ``present''.

\begin{table}
\begin{tabular}{|c|c|ccc|c|c|} \hline
&& \multicolumn{3}{|c|}{Counts out of $47\times233$} & Mean ``False & Mean ``False \\ \cline{3-5}
Model & AUC & Reversed & Tied & Correct & Positives'' & Negatives'' \\ \hline 
Read counts only (model-free) & 0.957 & 238 & 429 & 9814 & N/A & N/A \\
P-TG, independent samples & 0.932 & 583 & 259 & 9639 & 0.031 & 0.148 \\ 
P-TG, replicates recognized & 0.951 & 371 & 294 & 9816 & 0.031 & 0.064 \\
NB-NB, independent samples & 0.963 & 258 & 259 & 9964 & 0.100 & 0.050 \\
NB-NB, replicates recognized & 0.975 & 118 & 294 & $10{,}069$ & 0.092 & 0.030 \\ \hline
\end{tabular}
\caption{Area under curve and mean probabilities of false assignment 
for different models.  P-TG refers to the Poisson-Truncated Geometric model 
of \citet{Bao2014}, whereas NB-NB is the negative binomial-negative binomial model
introduced in this article.}
\label{AUCtable}
\end{table}

The fact that all four methods achieve AUC scores near 100\% indicates the relative ease of 
assignment for the particular six integration sites considered.  However, despite the high scores,
the striking differences in the performance of the methods is evident from the proportion 
of discordant pairs {\em not} correctly categorized, obtained by subtracting AUC from one.  By this 
measure, the Poisson-Truncated Geometric model of \citet{Bao2014} makes roughly 2.7 times as many
errors as the model we propose in this article.  In addition, the AUC scores indicate 
that recognizing animal replicates is important, as errors are reduced by almost a third by doing so.
Finally, we make the interesting and important observation that the model we propose in this
article performs better on these test data than the read count data themselves, which serve as
a sort of benchmark for an idealized thresholding method.  In fact, no thresholding method could likely
do this well, since some discordant pairs that are scored as correct using the read counts are likely
to fall entirely above or below whatever read count threshold is chosen, which would lower the number of
correct pairs.  At the same time, no discordant pair scored as reversed or tied according to the read 
counts could possibly become correct using a thresholding method, although it is true that some reversals
could change to ties.

In addition to AUC, Table~\ref{AUCtable} reports the mean probabilities of incorrect assignment for both
the 47 present sites and the 233 absent sites.  In other words, the mean ``false positive'' probability is
the mean value of $\hat Z_{ij}$ for the 233 absent sites, and the mean ``false negative'' probability is the mean
value of $1-\hat Z_{ij}$ for the 47 present sites.  We see that by this measure,  
the Poisson-truncated geometric models of \citet{Bao2014} do a better job for these 
particular data in the case where the sites are absent, but the
reverse is true when the sites are present.  Not surprisingly, both models are better when replicates are 
considered, since this uses more of the available information.

\subsection{Summarizing animal relationships}
\label{sec:animals}

There are many potential methods to analyze the probabilistic assignment of virus insertion sites---or,
more generally, TEs and alleles---to
animals represented by the $\hat Z$ matrix derived from our mixture
model and estimation procedure.  Broadly speaking, a suite of population genetics
tools exists to utilize allele frequency data to estimate population parameters.
Several of these methods accommodate probabilistic assignments as well.
As an example, \citet{Bao2014} demonstrate a hierarchical clustering method that uses 
such probabilistic assignments.

\begin{figure}[htb]
\begin{tabular}{cc}
\begin{minipage}{.45\textwidth}
\includegraphics[width=\textwidth]{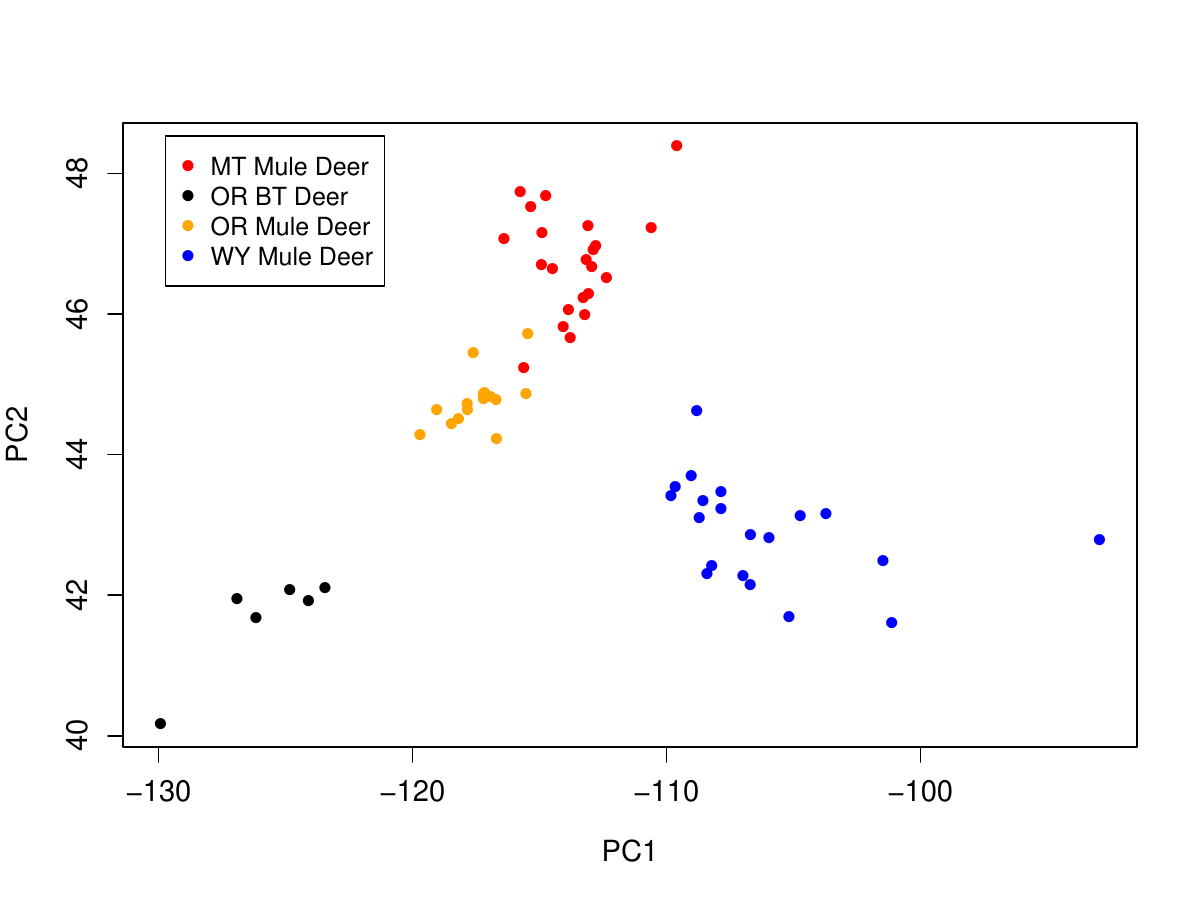}
\end{minipage}
&
\begin{minipage}{.45\textwidth}
\includegraphics[width=\textwidth]{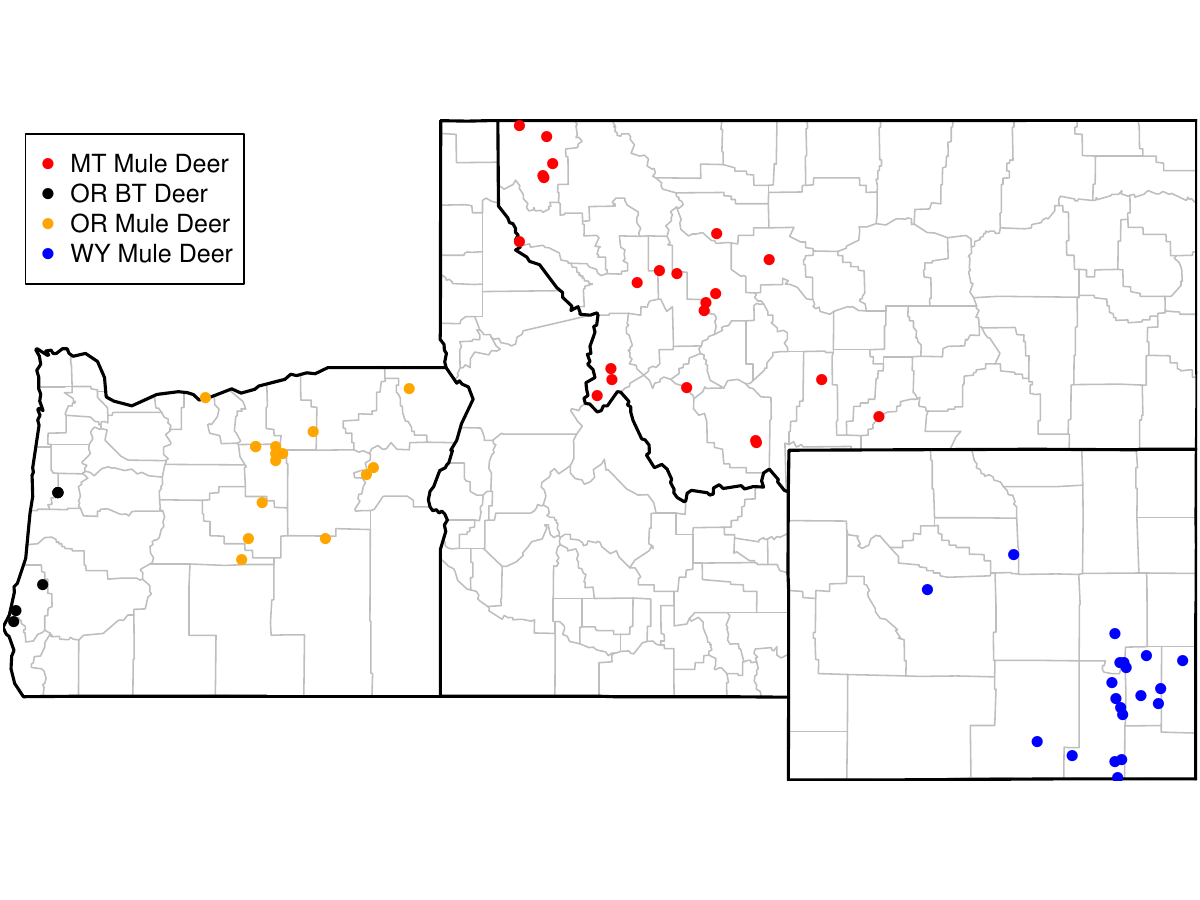}
\end{minipage}
\end{tabular}
\caption{\footnotesize{First two principal component scores (left) and geographic locations (right)
of the deer.  In the legends, BT stands for blacktail and MT, OR, and WY stand for Montana, 
Oregon, and Wyoming.}}
\label{fig:PCA}
\end{figure}

Here, we illustrate one type of analysis based on the information provided by the
$\hat Z$ matrix to estimate how variation in CrERV integration sites is distributed
among the animals from the four sampled populations.  
By considering each column of this matrix as a point in $m$-dimensional space, 
we may perform principal components analysis (PCA) and visualize the first two principal components.
In Figure~\ref{fig:PCA}, we see a depiction of the result after the first two PC scores are rotated and
scaled so as to make their two-dimensional locations comparable with the geographic locations 
where the animals were found.

The deer depicted separately from the others
in the lower left quadrant of Figure~\ref{fig:PCA} are the blacktail deer
subspecies of mule deer that emerged about
$20{,}000$ years ago.  The close association of Oregon and Montana mule deer to each other 
and the more distant relationship of Wyoming animals is an unexpected finding, given that previous
studies have reported low population subdivision in mule deer
\citep{Powell2013,Cullingham2011}.

\section{Discussion}
\label{sec:discussion}

The goal of our research was to determine which individuals share a genomic feature, in this case a newly 
described endogenous retrovirus, at a particular site in the genome. 
The data used to determine the presence of a polymorphic genome feature are often based 
on the number of reads assigned to it. Read count data are heavily skewed toward small numbers, 
creating uncertainty in the presence/absence status of any particular element.  Our article demonstrates 
the utility of using a mixture model to assign a probability that an insertion site is present in a given individual. 
Because these retroviruses are inherited like any host gene, animals that share more insertion sites are more 
closely related.  Our 
results show that animals from Wyoming can be distinguished from those from the adjacent state of Montana 
based on the profile of shared virus integration sites. This is a surprising finding because mule deer are 
migratory animals and can move between these two geographic locations. In fact, based on these analyses, 
the Montana mule deer appear more closely related to those in Oregon. Studies using traditional approaches 
report that mule deer have little population structure throughout this region 
\citep{latch2014,Powell2013,Cullingham2011}.

While we demonstrate the utility of using a mixture model for read count data for an endogenous retrovirus, 
our methodology is applicable to any data meant to determine the presence or absence of any polymorphic 
element---for instance, a different class of mobile element such as a long interspersed nuclear element, or
LINE \citep{akagi2008, evrony2012, burns2012, richardson2014}.  Indeed, these methods could apply 
beyond the biological realm to other situations in which data subjected to multiple sources of variability
include a large number of ``zeros'' that may not always be recorded as zeros as in the present application; 
the vast literature on zero-inflated models indicates that such applications could be myriad.

The primary statistical contributions of this article are twofold:  First, it reinforces and provides 
additional evidence
to support the argument made in \citet{Bao2014} that a two-component mixture model for
estimating probabilities of binary outcomes being positive, given observed count data,
is more flexible, principled, and accurate than the commonly-used approach of
dichotomizing results based on a count threshold.
Second, it significantly advances the mixture approach proposed by \citet{Bao2014} by 
carefully considering the statistical features of these data.  As one indication that
the fitted model gives sensible results, we find that in all cases, the best-fitting
parameters imply that $E (X_{ij} | \mbox{$j$ contains $i$}) > 
E (X_{ij} | \mbox{$j$ does not contain $i$})$ even though,
as explained in Section~\ref{sec:mixmod},
we do not enforce this inequality 
using constraints.

Our approach has the additional feature that it allows seamless integration of 
data from multiple batches.  This is prudent because 
not all samples included in an analysis are processed at the same time.
Experimental realities such as different ``absent'' count distributions for different
batches and samples that are replicated in more than one experiment can
be automatically accounted for by the model.  As a case in point, the read counts 
we analyze in this article are a superset of the counts used by 
\citet{Bao2014}.

In our dataset, the counts from multiple 
experiments all used the same Ion Torrent sequencing platform; yet in principle 
the model we propose can incorporate data from different platforms as well, which 
is important because sequencing technology advances rapidly and
so techniques such as ours that do not necessitate discarding ``old'' runs are both scientifically
prudent and economical.
Indeed, the adoption of our method enables the experimenter to consider designing
experiments that include some replicated animals between experiments since this overlap
will serve to validate the results.  This leads to further questions of how to design such 
experiments optimally to achieve the best tradeoff of statistical accuracy and experimental
cost, which could be considered in future work.

\section*{Acknowledgments}
We are grateful to the editor, associate editor, and two reviewers for numerous insightful comments that
led to substantial improvements.

\bibliographystyle{imsart-nameyear}
\bibliography{MixModLib}

\end{document}